\documentstyle[aps,prl,twocolumn,epsf,floats]{revtex}

\newcommand{\beq}{\begin{equation}}
\newcommand{\eeq}{\end{equation}}
\newcommand{\bea}{\begin{eqnarray}}
\newcommand{\eea}{\end{eqnarray}}

\newcommand{\etal}{{\it et al.},}

\begin{document}

\title{\bf \LARGE
Comment on "Structure of a Quantized Vortex near the BCS-BEC Crossover in an Atomic Fermi Gas"}
\vspace{0.75cm}
\author{ Aurel Bulgac}
\vspace{0.50cm}

\address{Department of Physics, University of
Washington, Seattle, WA 98195--1560, USA}

\maketitle

\begin{abstract}

A comment on the letter by M. Machida and T. Koyama, Phys. Rev. Lett. {\bf 94}, 140401 (2005) and also on the preprint by Y. Kawaguchi and T. Ohmi, cond-mat/0411018.

\end{abstract}

\draft


%
%
%
%


Machida and Koyama presented a study of the quantized vortex core structure near the BCS-BEC crossover regime \cite{machida}. A very similar analysis was performed by Kawaguchi and Ohmi \cite{ohmi}. The conclusions are similar, based on essentially the same theoretical approach due to Timmermans \etal \cite{eddy}. Initially  this theoretical approach was believed (incorrectly) to handle in a satisfactory manner the case of the large scattering length $a$, when $n|a|^3>1$, where $n$ is the atom number density. In this approach one introduces a boson degree of freedom, associated with two atoms forming a boson molecule in the closed channel. The practitioners of this approach,  conclude typically that in the BCS-BEC crossover regime there is a significant, even dominant, probability of the atoms to be in the closed channel. If that would be the case, then an atomic Fermi gas in this regime would behave essentially as a Bose system.  The probability of being in the closed channel is negligible \cite{abgfb} and a recent direct measurement of this quantity confirms this unequivocally  \cite{randy}.
The authors of Refs. \cite{machida,ohmi} seem to arrive at qualitatively similar conclusions to those obtained by the author and Y.Yu in Refs. \cite{abyy}, that in the BCS-BEC crossover regime a vortex in an atomic Fermi gas shows an unexpected \cite{bruun} marked 
density depletion. However, these authors argue, incorrectly, that this density depletion is due (mostly) to the significant, even dominant, occupation probability of the closed boson channel. 

On one hand, the two channel approach \cite{eddy} suffers from several deficiencies. While the physics is clearly determined by a single dimensionless parameter only, the typical one used being $1/k_Fa$, where $n=k_F^3/3\pi^2$, the model of Refs. \cite{machida,ohmi,eddy} is overdetermined
($U$ - the atom-atom "bare interaction," $g$ - the closed channel boson to two atom coupling, $\nu$ - the detuning of the closed channel, an ill defined energy cutoff, etc.). 
Moreover, the typical usage of this model is within meanfield (with some fluctuations included sometimes). It is well known that the corrections to the meanfield are controlled by the parameter $k_F|a|\gg 1$. Moreover, a certain type of fluctuations (which are routinely ignored in these treatments) lead to a strong reduction of the pairing field both in the weak \cite{gorkov} and strong coupling limits \cite{carlson}. The calculations of Refs. \cite{machida,ohmi} neglect the role of the attractive meanfield too, which thus disfavors a density depletion. More to the point, experiment \cite{randy} shows unequivocally that in the BCS-BEC crossover the boson component contributes $\approx 3\times 10^{-6}\cdots 2\times 10^{-4}$, as opposed to the theoretical predictions of Refs. \cite{machida,ohmi}, namely $\approx 0.4\cdots 1$. Clearly, such an insignificant (as observed) fraction of (composite) bosons cannot influence the vortex core structure.

On the other hand, the calculation of Refs. \cite{abyy} are based on a theoretically consistent extension of the DFT \cite{kohn} to fermionic superfluid systems \cite{slda} and on fully non-perturbative calculations of the homogeneous state \cite{carlson}, defined by one parameter only, namely $1/k_Fa$. These results show, that in spite of the quenching of the pairing gap due to fluctuations, and in the absence of any boson contribution (which would otherwise favor a density depletion), there is a significant density depletion at the vortex core, qualitatively consistent with the recent experimental observations \cite{martin}.



\end{document}